\documentclass[conference]{IEEEtran}
\ifCLASSINFOpdf
  \usepackage[pdftex]{graphicx}
\else
\fi
\usepackage{amsmath}
\usepackage{stfloats}
\usepackage{url}
\usepackage{amsmath}
\usepackage{mathtools}
\DeclarePairedDelimiter\abs{\lvert}{\rvert}

\begin{document}

\title{Product Offerings in Malicious Hacker Markets}

\author{
\IEEEauthorblockN{Ericsson Marin, Ahmad Diab and Paulo Shakarian} 
\IEEEauthorblockA{Arizona State University\\
Tempe, Arizona\\
\{ericsson.marin, ahmad.diab, shak\}@asu.edu} 
}

\maketitle

\begin{abstract}
Marketplaces specializing in malicious hacking products - including malware and exploits - have recently become more prominent on the darkweb and deepweb.  We scrape 17 such sites and collect information about such products in a unified database schema.  Using a combination of manual labeling and unsupervised clustering, we examine a corpus of products in order to understand their various categories and how they become specialized with respect to vendor and marketplace.  This initial study presents how we effectively employed unsupervised techniques to this data as well as the types of insights we gained on various categories of malicious hacking products.
\end{abstract}

\IEEEpeerreviewmaketitle

\section{Introduction}

Websites on the deepweb and darkweb specializing in the sale of malicious hacking products - such as malware platforms, software exploits and botnet rental - have become the venue of choice for online purchase of these items by cyber criminals.  In this paper, we leverage unsupervised learning to categorize and study the product offerings of 17 of these online markets.  Specifically, we describe how we used manual labeling combined with clustering techniques to identify product categories (Section~\ref{clustSec}), and then we analyze the results both quantitatively and qualitatively (Section~\ref{clInterp}).  We identify categories of products that are highly specialized with respect to particular vendors and markets.  We also highlight other interesting facets of this ecosystem - for instance, vendors who habitually cross-list products on multiple sites and nearly identical products for sale by multiple vendors.
\vspace{6pt}

\noindent\textbf{Background and Related Work.}  The darkweb refers to the anonymous communication provided by crypto-network tools such as "The Onion Router" (Tor), which is free software dedicated to protect the privacy of its users by obscuring traffic analysis as a form of network surveillance \cite{Dingledine:2004}.  On the other hand, the deepweb refers to sites not indexed by common search engines due to a variety of reasons (e.g. password protections), that not necessarily rely on additional protocols.  

The sites on the darkweb and deepweb explored in this study comprise marketplaces \cite{Vincenzo:2015}. In these websites, vendors advertise and sell their goods and services relating to malicious hacking, drugs, pornography, weapons and software services. Products are most often verified before any funds are released to the seller. The main engine of these environments is trust. If a seller is misleading or fails to deliver the appropriate item, he is banned from the site. Similarly, buyers can be banned for not complying with the transaction rules. Basically, all marketplaces in darkweb require a registration and a valid account to get access. Sometimes, this registration process is not trivial, including effort to answer questions, solve puzzles, mathematical equation or CAPTCHA.  

Most related work on darkweb markets such as \cite{Christin:2013} focus on a single market and do not restrict their study to malicious hacking products.  Our previous work on markets~\cite{Robertson:2016} focused on a game theoretic analysis of a small subset of the data in this paper - and did not attempt to categorize the products for sale.  Additionally, there is a complementary lines of work on malicious hacking forums (i.e. \cite{Yang:2011,Benjamin:2015,Macdonald:2015,Zhao:12,Chen:2011,Shakarian:2015,Shakarian:2016,shak16chap}) - which is a related but different topic from this paper.
\section{Malicious Hacking Product Categorization}
\label{clustSec}
In this section, we describe our malicious hacker product dataset and our use of clustering to identify malicious hacker product categories.  We examined 17 malicious hacker marketplaces crawled over a 6 month period.  The crawled information was then parsed and stored in a relational database.  Relevant tables record the marketplaces themselves, the vendors of the various products and the items/products for sale.  Each item is associated with a vendor and a marketplace, allowing for join queries. Some of the more relevant fields for marketplace items include the price, title, description, rating, posting date.  In this work, we primarily extract features from the product title/name to generate features.  

\begin{figure}[!ht]
\centering
\includegraphics[scale=0.25,keepaspectratio]{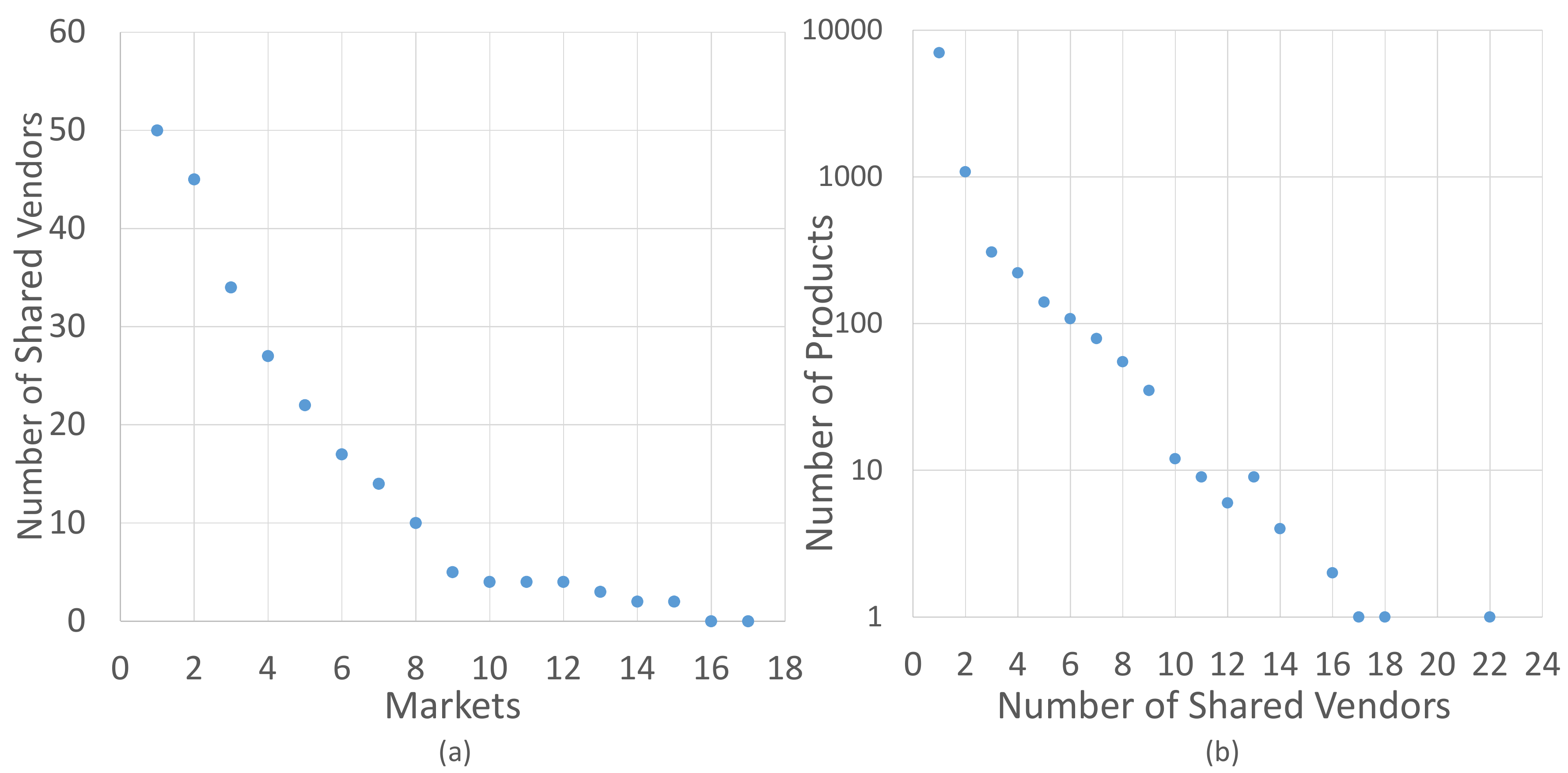}
 \caption{Distribution of (a) Shared Vendors over Markets. (b) Products over Shared Vendors. }
\label{fig:fig_distribution_linear}
\end{figure}

We note that many items are cross-posted and are nearly identical.  We show the distribution of vendors who use the same screen-name across multiple marketplaces in Fig. ~\ref{fig:fig_distribution_linear}(a).  To clean our product data, we identify duplicate (cross-posted) products and report on the size of our dataset in Table~\ref{tab:scraped_data}.  As we collect data from a variety of different sites,  there is inconsistency as to how products are categorized on each site - if such non-trivial categorization even exists for a given site.  In addition, there is a clear absence of a standardized method for vendors to register their products. As a consequence, the great majority of products are unique when compared with simple matching or regular expression technique. It is valid even in the case where a pair of vendors with different screen names post what a human would determine to be the same product. Fig. \ref{fig:fig_distribution_linear}(b) shows the distribution of products and number of vendors sharing each product.  The distribution follows a power-law.  Note that about 57\% of products are unique by simple comparison methods.

\begin{table}[!ht]
\centering
\caption{Scraped Data from Marketplaces in Darkweb.}
\begin{tabular}{|l|c|l|c|}

\hline  Marketplaces & 17 \\ 
\hline  Products (Total) & 16122 \\ 
\hline  Products (Distinct) & 9093 \\ 
\hline  Vendors & 1332 \\ 

\hline 

\end{tabular}
\label{tab:scraped_data}
\end{table}

\vspace{6pt}

\noindent\textbf{Clustering approach.} Using product names, we engineer features that represent each product as a vector. A set of pilot experiments suggests that word and character $n$-grams would provide more pure clusters compared with other feature engineering methods, such as meta-data or domain-specific keywords. These features were valued using standard term frequency - inverse document frequency (TF-IDF), after the elimination of stopping words and the execution of steaming.  We evaluated word $n$-gram feature vectors of length up to 1 and up to 2 words and many character $n$-gram features in the ranges from 3 to 7 and from 4 to 7. This gave us 10 different feature vectors in all. To verify which of these strategies could reach the best performance in our dataset, we evaluated the effect of the different types of feature vectors on the accuracy and purity of clusters produced by the K-means algorithm.

To determine the best feature vector, we manually labeled 500 samples using 34 labeled groups (listed in Table~\ref{tab: clusters_analysis_tab}).  We used 400 of the samples to determine centroids for each of the 34 groups, and then we evaluated the resulting clustering on the remaining 100 samples.  We examined the accuracy of the different approaches when compared to ground truth using the Rand-index method \cite{Rand:1971}. This method is defined as the number of pairs correctly considered in the same class or correctly considered in different classes divided by ${n \choose 2}$, where $n$ is the number of samples. In addition, we used standard entropy measurements to examine the purity of the clusters. Entropy measures the amount of disorder in a cluster. A zero-value for this metric means the clusters are formed by just one class.  The formal definition is as follows:

\begin{equation}
entropy(D_{i}) = - \sum_{i = 1}^{k} Pr_{i}(c_{j})\log_2 Pr_{i}(c_{j}),
\end{equation}

where $Pr_{i}(c_{j})$ is the proportion of class $c_{j}$ data points in cluster $i$ or $D_{i}$. The total entropy (considering all clusters) is:

\begin{equation}
entropy_{total}(D) = \sum_{i = 1}^{k} \frac{\abs{D_{i}}}{\abs{D}} ~\textit{x}~ entropy(D_{i})
\end{equation}

Table \ref{tab: clustering_fixed_evaluation} shows the performance of each TF-IDF vectorization using Rand-index and entropy, when K-means starts with the 34 fixed centroids. For Rand-index, character $n$-grams in the range from 3 to 4, 3 to 5, and 3 to 6, when K-means used cosine similarity reached a high best performance (0.986). In addition, we also found the best entropy (0.067) when K-means uses the same specification. This way, K-means configuration with character $n$-grams in the range from 3 to 6 for vectorization, cosine similarity for distance function and the 34 points for the starting centroids was our natural choice to produce the clusters in the entire dataset. 

We also examined the performance of our approach using random centroids. As expected, it performs worse than using the centroids derived from products.  Additionally, we examined products (from the full dataset) with a cosine similarity of less than $0.1$ from the calculated centroids.  There were 410 such distinct products (4.51\% of the dataset).  These were then manually examined and most were found to be irrelevant to our target domain - and we did not consider them further. 

\begin{table*}[!ht]
\centering
\scriptsize{
\caption{\centerline{K-means Evaluation (Fixed Centroids).}}
\begin{tabular}{|c|c|c|c|c|c|c|c|c|c|c|c|}

\hline \multicolumn{12}{|c|}{Rand-index} \\
\hline  & word(1,1) & word(1,2) & char(3,4) & char(3,5) & char(3,6) & char(3,7) & char(4,4) & char(4,5) & char(4,6) & char(4,7) & Random \\ 
\hline Cosine    & 0.986 & 0.985 & 0.986 & 0.986 & 0.986 & 0.985 & 0.985 & 0.985 & 0.984 & 0.982 & 0.933 \\ 
\hline Euclidean & 0.986 & 0.977 & 0.976 & 0.973 & 0.973 & 0.974 & 0.975 & 0.975 & 0.977 & 0.971 & 0.933  \\ 
\hline \multicolumn{12}{|c|}{Entropy} \\
\hline Cosine    & 0.075 & 0.079 & 0.067 & 0.067 & 0.067 & 0.075 & 0.072 & 0.079 & 0.088 & 0.088 & 0.423 \\ 
\hline Euclidean & 0.224 & 0.110 & 0.153 & 0.156 & 0.156 & 0.141 & 0.134 & 0.134 & 0.137 & 0.175 & 0.423  \\ 

\hline 

\end{tabular}
\label{tab: clustering_fixed_evaluation}
}
\end{table*}
\section{Analyst Interpretation of Product Clusters}
\label{clInterp}

In this section, we examine the results of clustering based on character $n$-grams in the range 3 to 6 using initial centroids determined from the labeled data.  In order to analyze the information of these clusters, we calculated their entropy with respect to two different criteria: marketplaces and vendors. We also checked in the database the number of distinct marketplaces and vendors inside each cluster. The idea was to understand the diversity of the clusters regarding these two facets. A low marketplace entropy for a given cluster would mean its products were mainly found in a particular marketplace. Similarly, low vendor entropy would mean the cluster's products were mainly sold by a particular vendor. Table \ref{tab: clusters_analysis_tab} presents the results.

\begin{table}[!ht]
\centering
\tiny{
\caption{Clusters Entropy.}}
\begin{tabular}{|c|l|c|c|c|c|c|}

\hline  & & N\textsuperscript{\underline{o}} of & N\textsuperscript{\underline{o}} of & Market & N\textsuperscript{\underline{o}} of & Vendor \\
 Rank & Cluster Name & Products & Markets & Entropy & Vendors & Entropy \\

\hline 1 & Carding & 1263 & 16 & 0.320 & 315 & 0.720 \\ 
\hline 2 & PayPal-related & 1103 & 16 & 0.340 & 335 & 0.754 \\ 
\hline 3 & Cashing Credit Cards & 867 & 16 & 0.351 & 256 & 0.738 \\ 
\hline 4 & PGP & 865 & 15 & 0.347 & 203 & 0.696 \\ 
\hline 5 & Netflix-related & 846 & 14 & 0.270 & 351 & 0.805 \\ 
\hline 6 & Hacking Tools - General & 825 & 15 & 0.331 & 132 & 0.516 \\ 
\hline 7 & Dumps - General & 749 & 12 & 0.289 & 280 & 0.777 \\ 
\hline 8 & Linux-related & 561 & 16 & 0.372 & 117 & 0.758 \\ 
\hline 9 & Email Hacking Tools & 547 & 13 & 0.335 & 196 & 0.738 \\ 
\hline 10 & Network Security Tools & 539 & 15 & 0.366 & 117 & 0.621 \\ 
\hline 11 & Ebay-related & 472 & 15 & 0.385 & 163 & 0.772 \\ 
\hline 12 & Amazon-related & 456 & 16 & 0.391 & 197 & 0.825 \\ 
\hline 13 & Bitcoin & 443 & 15 & 0.360 & 201 & 0.823 \\ 
\hline 14 & Links (Lists) & 422 & 12 & 0.211 & 221 & 0.838 \\ 
\hline 15 & Banking & 384 & 13 & 0.349 & 186 & 0.840 \\ 
\hline 16 & Point of Sale & 375 & 15 & 0.384 & 181 & 0.841 \\ 
\hline 17 & VPN & 272 & 12 & 0.413 & 130 & 0.827 \\ 
\hline 18 & Botnet & 257 & 12 & 0.291 & 110 & 0.796 \\ 
\hline 19 & Hacking Groups Invitation & 251 & 14 & 0.387 & 143 & 0.865 \\ 
\hline 20 & RATs & 249 & 15 & 0.453 & 99 & 0.797 \\ 
\hline 21 & Browser-related & 249 & 12 & 0.380 & 134 & 0.857 \\ 
\hline 22 & Physical Layer Hacking & 237 & 13 & 0.408 & 122 & 0.856 \\ 
\hline 23 & Password Cracking & 230 & 13 & 0.434 & 100 & 0.781 \\ 
\hline 24 & Smartphone - General& 223 & 14 & 0.408 & 110 & 0.816 \\ 
\hline 25 & Wireless Hacking & 222 & 13 & 0.389 & 56 & 0.601 \\ 
\hline 26 & Phishing & 218 & 13 & 0.403 & 111 & 0.849 \\ 
\hline 27 & Exploit Kits & 218 & 14 & 0.413 & 91 & 0.795 \\ 
\hline 28 & Viruses/Counter AntiVirus & 210 & 14 & 0.413 & 60 & 0.684 \\ 
\hline 29 & Network Layer Hacking & 205 & 14 & 0.459 & 60 & 0.716 \\ 
\hline 30 & RDP Servers & 191 & 12 & 0.405 & 124 & 0.895 \\ 
\hline 31 & Android-related & 156 & 11 & 0.429 & 60 & 0.770 \\ 
\hline 32 & Keyloggers & 143 & 13 & 0.496 & 77 & 0.862 \\ 
\hline 33 & Windows-related & 119 & 12 & 0.464 & 50 & 0.717 \\ 
\hline 34 & Facebook-related & 119 & 15 & 0.501 & 67 & 0.876 \\ 

\hline 

\end{tabular}
\label{tab: clusters_analysis_tab}
\end{table}

As shown in Table \ref{tab: clusters_analysis_tab}, \textit{Links} holds the lowest entropy when we analyzed the marketplaces, suggesting the great majority of products come from the same market. In this cluster, 80\% of products came from only 2 markets. However, when we check the vendor entropy for this same cluster, we can observe a higher value, suggesting that many vendors are actually offering products related to \textit{Links}.  It is possible that many markets discourage the re-selling of lists of links, as much of this information can be found on darkweb Wiki's for free.

Similarly, \textit{Hacking Tools} holds the lowest entropy for the vendor criteria. This suggest that only a few vendors are present in that cluster.  Specifically, only $2$ vendors author $416 (50\%)$ of this type of products.  At first glance, this may be surprising as this appears to be a very general group.  However, upon inspection of the contents, we find that many authors of these products are actually organizations.  These organizations use similar language in their product description in an effort to brand their wares.  This could indicate the presence of hacking-collectives that author products as well as the limitations of our text-based approach - which can potentially cluster products branded  in a similar fashion.  We also note one of the most prominent vendor in this  cluster was itself a marketplace - which is also reflected in the low marketplace entropy.

In our analysis of the \textit{Facebook} and \textit{Keylogger} clusters, we can see that they point to the other direction. They have high values for both entropy, a clear sign about the diversity with respect to both vendors and markets. For example, in cluster \textit{Facebook}, there were 119 products and 67 vendors, and the most prolific vendor for this cluster authored only 8 products. In the same cluster, products were also spread across 15 markets - and the most well-represented market was associated with 30 products.  This analysis also indicates widespread prevalence of keyloggers - which is not surprising as it is a well established hacking technique.  However, observing the similar trend for the \textit{Facebook} cluster could be indicative of an increase in demand for Facebook-directed social media hacking products and information.

\vspace{6pt}

\noindent\textbf{Conclusion and Future Work.}  In this paper, we conducted an initial examination of malware products from 17 malicious hacker markets through unsupervised learning.  Using manually-labeled data, we studied the effect of feature vector on cluster purity using text-based features.  We then analyzed the impurity of clusters in our corpus of over $8,000$ malicious hacking products with respect to vendor and marketplace, and finally, we identified several interesting characteristics of how the products were grouped.  Currently, we are examining other methods for grouping these products using matrix factorization and supervised techniques.  Additionally, we are studying the underlying social network of vendors through relationships based on similar product offerings.

\vspace{6pt}

\begin{small}
\noindent\textbf{Acknowledgments.} Some of the authors were supported by the Office of Naval Research (ONR) Neptune program, the Arizona State University Global Security Initiative (ASU GSI), and CNPq-Brazil.
\end{small}

\bibliographystyle{IEEEtran}
\bibliography{IEEEabrv,bib}

\end{document}